\documentclass[12pt,preprint]{aastex}

\slugcomment{}

\shorttitle{Helical motion of magnetic flux tubes}
\shortauthors{Zaqarashvili et al.}

\begin{document}


\title{Helical motion of magnetic flux tubes in the solar atmosphere}


\author{T.V. Zaqarashvili and N. Skhirtladze}
\affil{Abastumani Astrophysical Observatory at I. Chavchavadze State
University, Kazbegi Ave. 2a, Tbilisi 0160, Georgia}
\email{temury@genao.org, skhirtladze@genao.org}



\begin{abstract}
Photospheric granulation may excite transverse kink pulses in
anchored vertical magnetic flux tubes. The pulses propagate upwards
along the tubes with the kink speed, while oscillating wakes are
formed behind the wave front in a stratified atmosphere. The wakes
oscillate at the kink cut-off frequency of stratified medium and
gradually decay in time. When two or more consecutive kink pulses
with different polarizations propagate in the same thin tube, then
the wakes corresponding to different pulses may superimpose. The
superposition sets up helical motions of magnetic flux tubes in the
photosphere/chromosphere as seen by recent Hinode movies. The energy
carried by the pulses is enough to heat the solar chrmosphere/corona
and accelerate the solar wind.

\end{abstract}





\section{Introduction}

Recent high resolution movies obtained by Hinode spacecraft clearly
show continuous helical motions of spicule axes \citep{dep}. This
phenomenon is known for a long time (Beckers 1972), but no
satisfactory explanation has been done yet. \citet{dep} suggested
that these motions are caused by Alfv\'en waves excited in the
photosphere by granular motions or acoustic oscillations.
Photospheric magnetic field is concentrated in thin flux tubes and
therefore may support the propagation of kink and torsional Alfv\'en
waves (the tubes also support the propagation of sausage waves, but
here we consider only transverse waves). The kink wave is a tube
wave i.e. the whole tube oscillates with one frequency even if the
Alfv\'en speed changes across the tube, while the frequency of
torsional waves is different at different surfaces \citep{van}.
Therefore, the excitation of torsional waves in photospheric
magnetic tubes is complicated. The oscillations of spicule axis have
also been explained in terms of kink waves \citep{kuk,zaq}, but the
chromospheric magnetic field has rather expanded structure.
Therefore, \citet{dep} suggested through Monte-carlo simulation that
spicules are not wave guides for kink waves. The problem is
currently under debate and more observations are needed to
understand the real process.

On the other hand, granular buffeting on an anchored magnetic tube
may easily excite a transverse kink pulse. The pulse propagates
through the stratified photosphere with the kink speed, but the
oscillating wake is formed behind the wave front. The wake
oscillates at cut-off frequency of kink waves and decays as time
progresses \citep{rob1,rob,rae,spruit,has,mus}. However, the
magnetic tube undergoes continuous buffeting of granular cells from
different sides. Therefore, another pulse polarized in a different
plane may quickly follow. The second pulse again propagates with the
kink speed and form another wake oscillating with the same cut-off
frequency. Therefore, the superposition of the two oscillations may
set up the {\it helical motion} of the tube axis in the photosphere.
The helical motion will have the photospheric kink cut-off period
$\sim$ 7-8 min.

However, when the pulse penetrates into the chromosphere, then two
possible scenarios can be developed. If chromospheric magnetic field
has a tube structure, then the pulse continues to propagate as the
kink one. But if the magnetic field is not concentrated in tubes,
then it will be transformed into the Alfv\'enic pulse. Yet the pulse
will have the same main properties in both cases: it will propagate
at either the kink or Alfv\'en speed and the wake oscillating at the
chromospheric cut-off frequency will be formed behind the pulse. The
superposition of the wakes corresponding to different pulses may set
up helical motions of magnetic field lines just as in the
photospheric case. These helical motions may be responsible for the
oscillations of spicule axes as seen by Hinode movies.

Here we study the phenomenon using the Klein-Gordon equation for
wave propagation in the stratified atmosphere.

\section{Helical kink waves in the photosphere}

Kink wave propagation along vertical thin magnetic flux tube
embedded in the stratified field-free atmosphere is governed by the
Klein-Gordon equation \citep{rae,spruit,rob}
\begin{eqnarray}
{{\partial^2 Q}\over {\partial z^2}}-{{1}\over c^2_k}{{\partial^2
Q}\over {\partial t^2}} - {{\Omega^2_k}\over c^2_k} Q =0,
\end{eqnarray}
where $Q=\xi(z,t)\exp{(-z/4\Lambda)}$,
$c_k=B_0/\sqrt{4\pi(\rho_0+\rho_e)}$ is the kink speed, $\Lambda$ is
the density scale height and $\Omega_k=c_k/4\Lambda$ is the
gravitational cut-off frequency for isothermal atmosphere
(temperature inside and outside the tube is assumed to be the same
and homogeneous). Here $\xi(z,t)$ is the transversal displacement of
the tube, $B_0(z)$ is the tube magnetic field, $\rho_0(z)$ and
$\rho_e(z)$ are the plasma densities inside and outside the tube
respectively (the magnetic field and densities are functions of $z$,
while the kink speed $c_k$ is constant in the isothermal
atmosphere).

Eq. (1) yields simple harmonic solutions $\exp[{i(\omega t \pm k_z
z)}]$ with the dispersion relation
\begin{eqnarray}
\omega^2 - \Omega^2_k= c^2_k k^2_z,
\end{eqnarray}
where $\omega$ is the wave frequency and $k_z$ is the wave number.
The dispersion relation shows that the waves with higher frequency
than $\Omega_k$ may propagate in the tube, while the lower frequency
waves are evanescent.

Kink waves cause the transverse displacement of whole tube. The
displacement of tube in a simple harmonic kink wave is polarized
arbitrarily and the polarization plane depends on the excitation
source. Then the superposition of two or more kink waves polarized
in different planes may give rise to the complex motion of the tube.
The process is similar to the superposition of two plane
electromagnetic waves, where the waves with the same amplitudes lead
to the circular polarization, while the waves with different
amplitudes lead to the elliptical polarization. Consider, for
example, two harmonic kink waves with the same frequency but
polarized in $xz$ and $yz$ planes: $A_x=A_{x0}\cos(\omega t + k_z
z)$ and $A_y=A_{y0}\sin(\omega t + k_z z)$. The superposition of
these waves sets up the {\it helical wave} with circular
polarization if $A_{x0}=A_{y0}$. As a result, the tube axis rotates
around the vertical, while the displacement remains constant (Fig.
1). If $A_{x0} \not= A_{y0}$ then the resulting wave is elliptically
polarized. The superposition of few harmonics with different
frequencies and polarizations may lead to more complex motion of
tube axis.

However, simple harmonic kink waves hardly be excited in the
photosphere. The more realistic process is the impulsive buffeting
of granules on an anchored magnetic flux tube. For the sake of
simplicity, we consider the simplest impulsive forcing in both time
and coordinate. Then Eq. (1) looks as
\begin{eqnarray}
{{\partial^2 Q}\over {\partial z^2}}-{{1}\over c^2_k}{{\partial^2
Q}\over {\partial t^2}} - {{\Omega^2_k}\over c^2_k} Q =-
A_0\delta(t)\delta(z),
\end{eqnarray}
where $z>-\infty$, $t>0$, $A_0$ is constant and the pulse is set at
$t=0$, $z=0$.

The solution of this equation is the Green function for the
Klein-Gordon equation, which can be written as
\citep{lamb1,lamb2,duf}
\begin{eqnarray}
Q={{c_kA_0}\over {2}}J_0\left [\Omega_k\sqrt{t^2 - {{z^2}\over
{c^2_k}}} \right ]H\left [\Omega_k \left ( t- {{z}\over {c_k}}
\right )\right ],
\end{eqnarray}
where $J_0$ and $H$ are Bessel and Heaviside functions respectively.
Eq. (4) shows that the wave front propagates with the kink speed
$c_k$, while the wake oscillating at the cut-off frequency
$\Omega_k$ is formed behind the wave front and it decays as time
progresses \citep{rae,spruit,has,rob}. Fig. 2 shows the plot of
transverse displacement $\xi(z,t)=Q(z,t)\exp{(z/4\Lambda)}$, where
$Q$ is expressed by Eq. (4). The rapid propagation of the pulse is
seen, which is followed by the oscillating wake (the time is
normalized by the cut-off period $T_k=2\pi/\Omega_k$). Just after
the propagation of the pulse, the tube begins to oscillate with the
cut-off period at each height. The amplitudes of pulse and wake
increase upwards due to the density reduction, but the oscillations
at each height decay in time.

Hence, the transverse impulsive action on the magnetic tube at $t=0$
moment near the base of photosphere (set at $z=0$) excites the
upward propagating kink pulse, while the tube in the photosphere
oscillates at the photospheric kink cut-off frequency, $\Omega_k$,
which depends on the plasma $\beta$ parameter ($=8\pi p_0/B^2_0$)
inside the tube. In the case of temperature balance inside and
outside the tube, the kink speed can be expressed as $c_k=c_s[\gamma
(1+2\beta)/2]^{-1/2}$, where $c_s$ is the sound speed and $\gamma$
is the ratio of specific heats ($\gamma=5/3$ for adiabatic process).
Then the photospheric sound speed of 7.5 km/s and $\beta=0.3$ gives
6.5 km/s for the kink speed. Consequently, we may estimate the kink
cut-off period as $\sim$ 8 min using the photospheric scale height
of 125 km. Hence the magnetic tube will oscillate with $\sim$ 8 min
period in the photosphere. If the external pulse is directed along,
say, the $x$ axis, then the tube will oscillate in the $xz$ plane.

However, the anchored magnetic tube undergoes the granular buffeting
from different sides. Therefore, suppose that after $t_0$ time
another granular cell acts on the same tube along the $y$ axis. The
solution governing the pulse propagation is
\begin{eqnarray}
\xi_y={{c_kA_0}\over {2}}J_0\left [\Omega_k\sqrt{(t-t_0)^2 -
{{z^2}\over {c^2_k}}} \right ]H\left [\Omega_k \left ( t-t_0-
{{z}\over {c_k}} \right )\right ]\exp{(z/4\Lambda)}.
\end{eqnarray}
Thus the rapidly propagating pulse is again excited with the
oscillating wake behind the front. The wake oscillates with the same
cut-off frequency, but the oscillation is polarized in $yz$ plane.
Hence there are two transverse oscillations with the same frequency,
but polarized in perpendicular planes. The time interval between
consecutive buffeting $t_0$ (say, granular life time) is comparable
to the photospheric kink cut-off period. Therefore, these
oscillations will be superimposed, because the oscillation excited
by the previous pulse still exists in the same tube. The
superposition will set up the helical motion of the tube axis with
photospheric cut-off period $\sim$ 8 min. Figure 3 shows the
superposition of the solutions (4) and (5) at the height of 250 km
above the photosphere. The first pulse is imposed along the $x$
direction, which is followed by another pulse in the $y$ direction.
The upper panel corresponds to the same amplitudes of both pulses,
but the lower panel corresponds to the case when the first pulse is
twice stronger than the second. We see that the tube rotates along
nearly circular spiral in the first case and along elliptical spiral
in the second case. The displacement gradually decreases with time.
Therefore, the granular buffeting with same amplitudes excites the
nearly circular motion of the tube, while the buffeting with
different amplitudes excites the elliptical motion.

The wave length of oscillations $\lambda_c=c_kT_k \sim 3000$ km is
quiet long comparing to the width of the photosphere. Therefore, the
photospheric magnetic tube will just rotate around the vertical
without additional wave nodes. The tube displacement increases with
height due to the decreasing density (Fig. 2). Thus the observations
should show that the upper part of the tube rotates with larger
amplitude than the lower part. We believe that the high resolution
observations will reveal the similar behavior of photospheric
magnetic tubes.

\section{Propagation of transverse pulse through the chromosphere}

The situation is changed when the pulse crosses the photosphere and
penetrates into the chromosphere. Photospheric magnetic tubes may
expand in the chromosphere giving rather different geometry than
thin tubes. On the other hand, chromospheric spicules seem to behave
like magnetic tubes. But recent Monte-carlo simulations (De Pontieu
et al. 2007) suggest that the spicules are not wave guides for tube
waves. Therefore, this question is currently under debate and more
observations are needed to clarify the intrinsic process (Erd\'elyi
\& Fedun 2007). The transverse pulse retains its properties in any
case. It continues to be the kink pulse in structured magnetic
field, but probably is transformed into the Alfv\'enic one in the
case of smooth transverse profile of the magnetic field.

The photosphere and the chromosphere can be approximated as two
different regions with different isothermal temperatures, densities
and other plasma parameters. Then the propagation of pulse in the
chromosphere is governed by Eq. (3), but with different phase speed
and scale height ($z=0$ now corresponds to the base of the
chromosphere). It must be mentioned, however, that the phase speed
remains constant only if the magnetic field is expanded with height.
This necessarily requires the horizontal component of the magnetic
field, which is neglected in the equation. Therefore, the equation
(3) is valid only near the tube axis, where the magnetic field is
predominantly vertical.

Then the photospheric solution (4) can be directly applied here, but
with chromospheric phase speed and scale height. The chromospheric
scale height $\Lambda_{ch}$ can be estimated as $\sim$ 500 km for 25
000 K temperature. The value of phase speed determines the wave
cut-off frequency. For example, Alfv\'en wave cut-off frequency is
$\Omega_{cA}=v_A/4\Lambda_{ch}$ \citep{rob}, which gives the cut-off
period of $\sim$ 250 s for the Alfv\'en speed of 50 km/s. On the
other hand, a cavity with higher density concentrations (for example
spicules) may guide kink waves with smaller phase speed. This
increases the cut-off period. For example, the kink speed of 25 km/s
yields the cut-off period of 500 s.

Therefore, the transverse pulse may set up the oscillating wake in
the chromosphere with the period of 250-500 s. The two
perpendicularly polarized transverse pulses may form the helical
motion in the chromosphere as observed by Hinode (De Pontieu et al.
2007). De Pontieu et al. (2007) argued that the helical motion is
caused by Alfv\'en waves directly excited in the photosphere. The
estimated energy flux of the waves was enough to power the solar
wind and to heat the quiet corona. However, if observed oscillations
of spicule axes are caused by wakes formed after the transverse
pulse propagation, then the energy transported into the
chromosphere/corona can be much higher: {\it as almost whole energy
of initial perturbation is carried by the pulse, while the energy of
the wake is much smaller}.

The energy flux stored in initial transverse pulse at the
photospheric level is $F \sim n_e c_k v^2_g$, where $v_g$ is the
granular velocity being 1-2 km/s. Then for photospheric values of
electron density and kink speed, the estimated energy flux is $\sim
5{\cdot}10^8$ erg cm$^{-2}$ s$^{-1}$. Almost whole energy is carried
by the pulse, therefore even if the filling factor of magnetic tubes
is $10\%$, the energy flux is more than enough to heat the solar
chromosphere/corona.

\section{Conclusions}

We suggest that the propagation of consecutive transverse pulses in
the stratified atmosphere, which are excited by photospheric
granular buffeting, may set up the helical motions of magnetic flux
tubes through the superposition of oscillating wakes formed behind
the wave fronts. This scenario may explain the continuous motions of
spicule axes seen in recent Hinode movies. The pulses carry almost
whole energy of initial perturbations, while the energy in wake
oscillations is much smaller. Therefore, the energy carried into
corona by transverse pulses can be much higher than it is estimated
by observed oscillations. More observations and numerical/analytical
works need to look further into this problem.

\acknowledgments

The work was supported by the grant of Georgian National Science
Foundation GNSF/ST06/4-098.

\clearpage




\begin{figure}
\includegraphics[angle=270,scale=0.95]{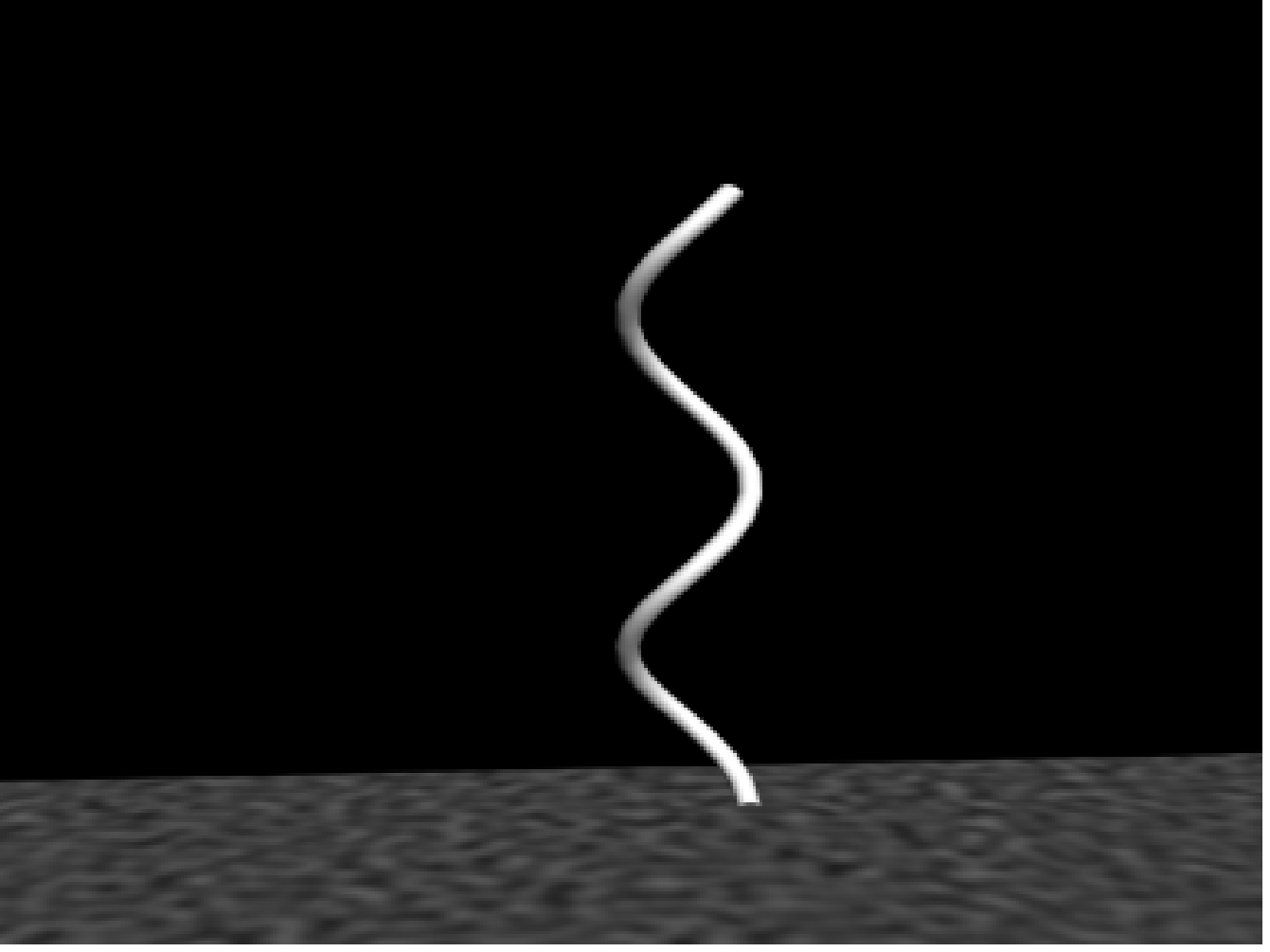} \caption{Helical kink wave in a thin magnetic flux tube.}
\end{figure}

\begin{figure}
\includegraphics[angle=0,scale=0.95]{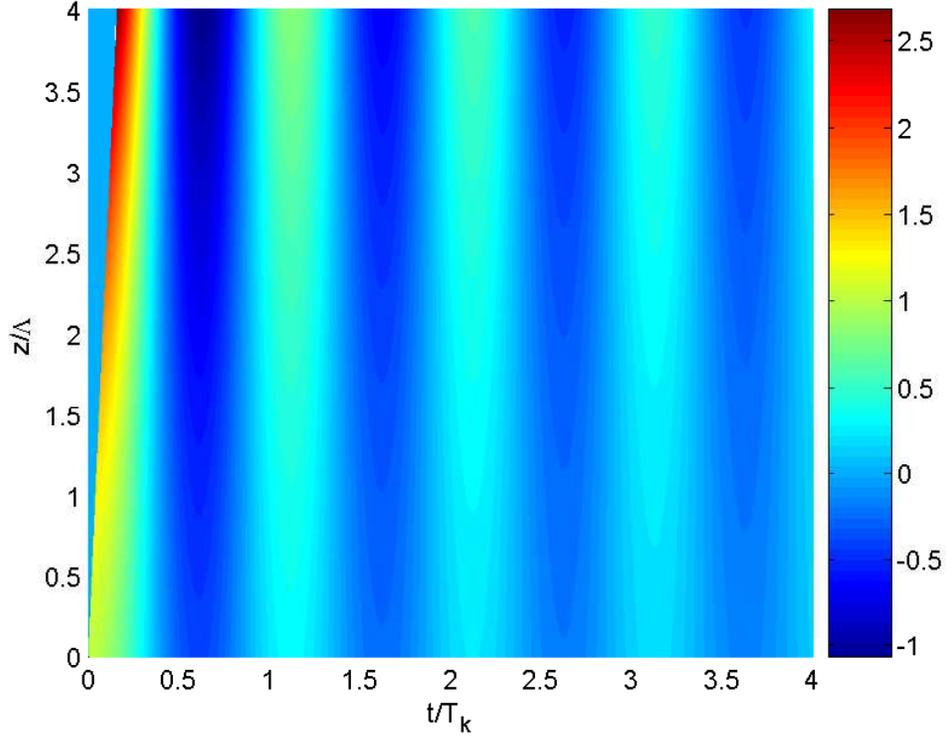} \caption{Propagation of transverse kink pulse in photospheric magnetic flux tubes.
Here we plot the tube transverse displacement
$\xi(z,t)=Q(z,t)\exp{(z/4\Lambda)}$ as a function of time $t$ and
height $z$, where $Q$ is expressed by Eq. (4). The time is
normalized by kink cut-off period and the $z$ coordinate is
normalized by the photospheric scale height, $\Lambda=125$ km. The
rapid propagation of the pulse is seen, which is followed by the
oscillating wake. The pulse propagates with the kink speed and the
wake oscillates with the cut-off period. The amplitude of the pulse
(and wake) increases with height due to the decreasing density. }
\end{figure}

\begin{figure}
\includegraphics[angle=0,scale=0.9]{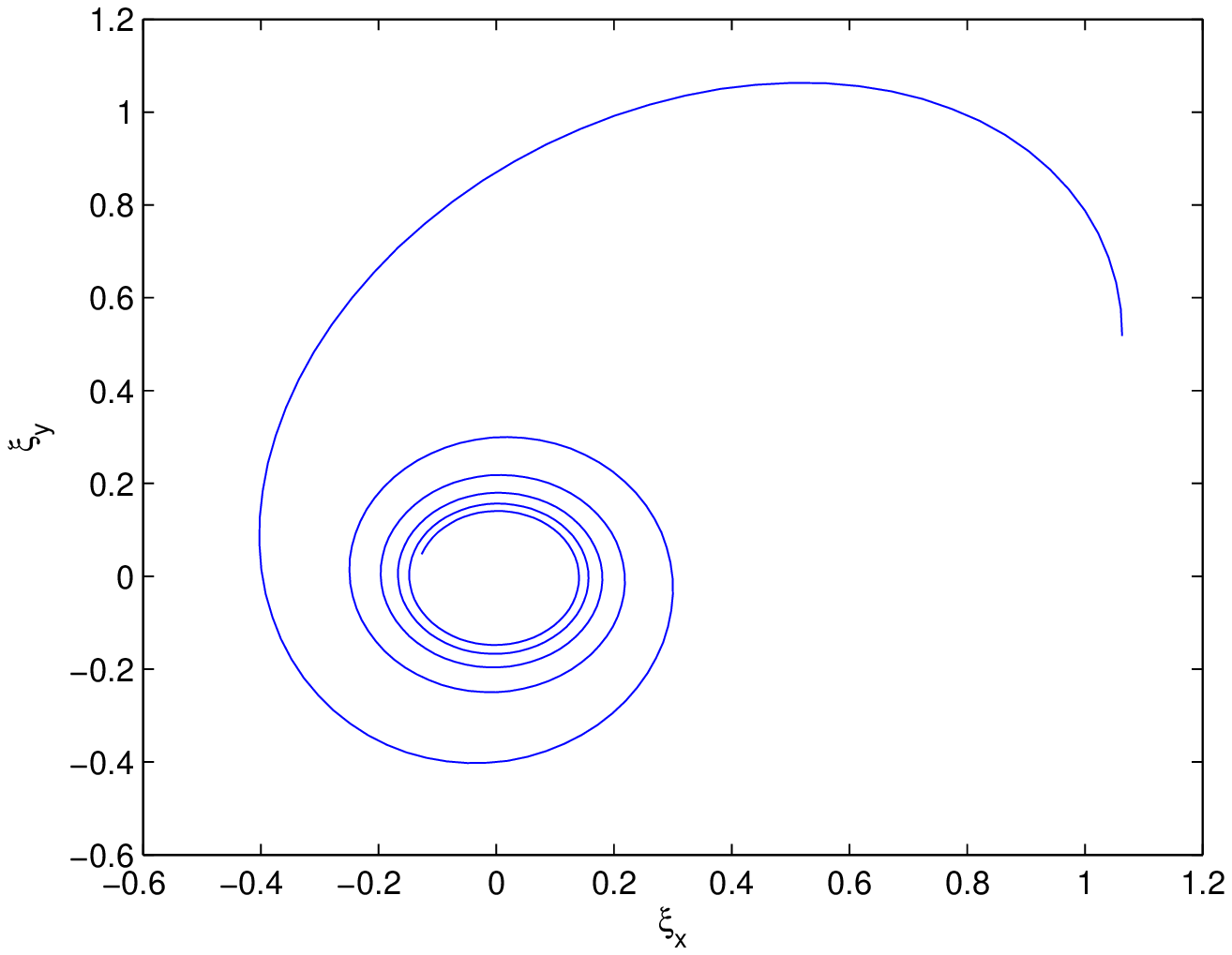}
\includegraphics[angle=0,scale=0.9]{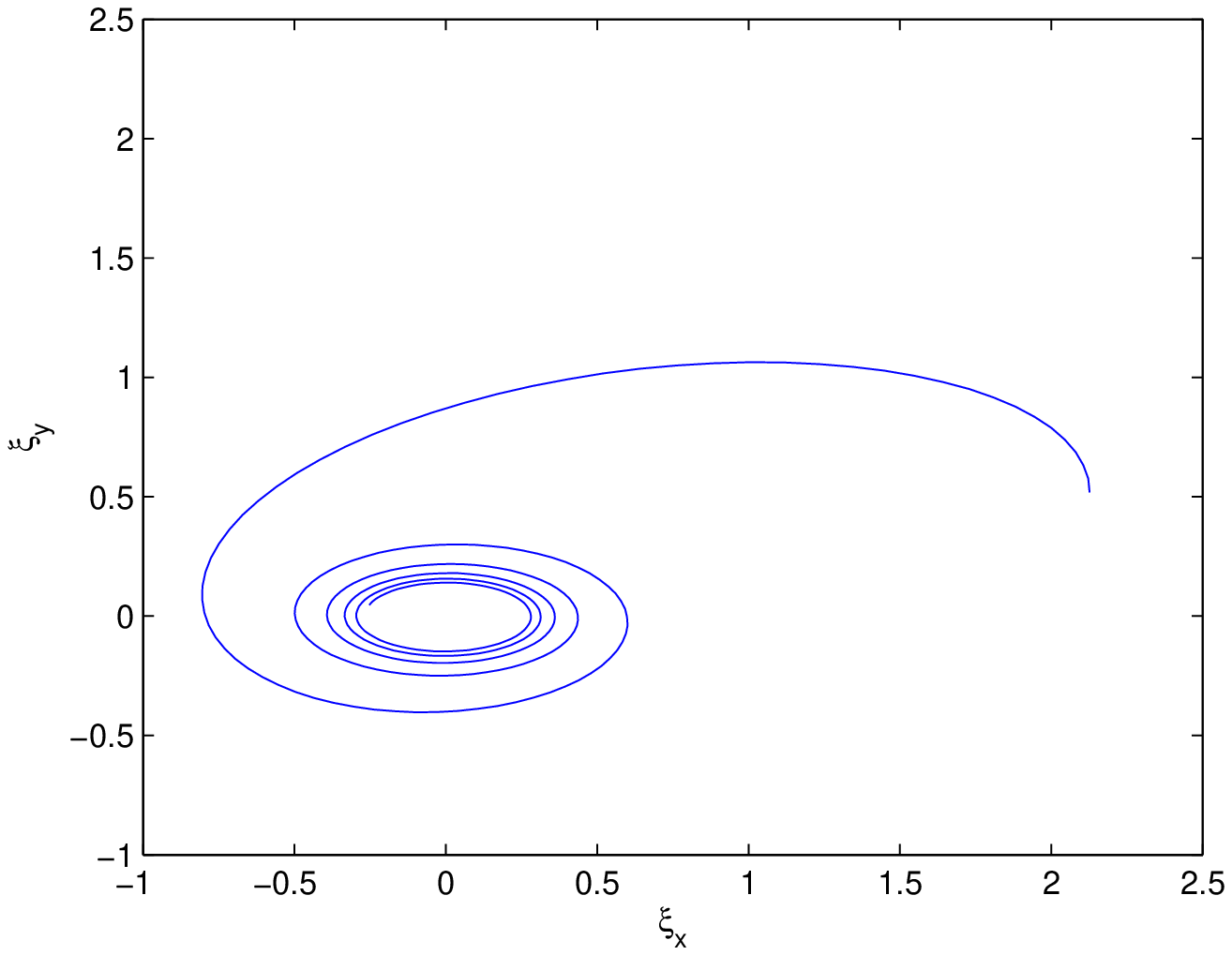}
 \caption{The helical motion of the tube axis at the height of 250 km from the surface due to the propagation of two consecutive
 transverse pulses polarized in $xz$ and $yz$ planes.
 Upper panel corresponds to the case of same amplitudes, while the lower panel corresponds to the case of different amplitudes.
 The pulses with the same amplitudes set up the nearly circular motion of the tube axis. While the pulses with different amplitudes lead to the
 elliptical motion of the tube axis. The displacement of tube axis gradually decreases with time.}
\end{figure}








\end{document}